\documentclass[preprint,11pt]{article}
\usepackage{color}
\usepackage{graphicx}
\usepackage[dvips]{epsfig}
\usepackage{graphics} 
\usepackage{times} 
\usepackage[cmex10]{amsmath} 
\usepackage{amssymb}  
\usepackage{cite}
\usepackage{multirow}
\usepackage[tight,footnotesize]{subfigure}
\usepackage{hyperref}

\def\twon #1{\left\|#1\right\|_2}
\def\onen #1{\left\|#1\right\|_1}

\def\frobn #1{\left\|#1\right\|_{\text{F}}}

\def\abs #1{\left|#1\right|}

\def\bC{\mathbb{C}}

\def\bR{\mathbb{R}}

\def\m #1{\boldsymbol{#1}}

\def\cU{\mathcal{U}}

\def\ccn{\mathcal{CN}}

\def\bee{\begin{equation}}
\def\ene{\end{equation}}

\def\beq{\begin{eqnarray}}
\def\enq{\end{eqnarray}}
\def\lentwo{\setlength\arraycolsep{2pt}}

\newtheorem{rem}{Remark}

\def\equ #1{\begin{equation}#1\end{equation}}
\def\equa #1{\begin{eqnarray}#1\end{eqnarray}}
\def\sbra #1{\left(#1\right)}
\def\mbra #1{\left[#1\right]}
\def\lbra #1{\left\{#1\right\}}
\def\diag #1{\text{diag}#1}

\title{Off-grid Direction of Arrival Estimation Using Sparse Bayesian Inference}
\author{Zai Yang, Lihua Xie\thanks{Z. Yang and L. Xie are with EXQUISITUS, Centre for E-City, School of Electrical and Electronic Engineering, Nanyang Technological University, 639798, Singapore (e-mail: yang0248@e.ntu.edu.sg; elhxie@ntu.edu.sg).}, \emph{Fellow, IEEE},
and Cishen Zhang\thanks{C. Zhang is with the Faculty of Engineering and Industrial Sciences, Swinburne University of Technology, Hawthorn VIC 3122, Australia (e-mail: cishenzhang@swin.edu.au).}}
\begin{document}

\date{}
\maketitle

\begin{abstract}
Direction of arrival (DOA) estimation is a classical problem in signal processing with many practical applications. Its research has recently been advanced owing to the development of methods based on sparse signal reconstruction. While these methods have shown advantages over conventional ones, there are still difficulties in practical situations where true DOAs are not on the discretized sampling grid. To deal with such an off-grid DOA estimation problem, this paper studies an off-grid model that takes into account effects of the off-grid DOAs and has a smaller modeling error. An iterative algorithm is developed based on the off-grid model from a Bayesian perspective while joint sparsity among different snapshots is exploited by assuming a Laplace prior for signals at all snapshots. The new approach applies to both single snapshot and multi-snapshot cases. Numerical simulations show that the proposed algorithm has improved accuracy in terms of mean squared estimation error. The algorithm can maintain high estimation accuracy even under a very coarse sampling grid.

\end{abstract}


\section{Introduction}


Source localization using sensor arrays \cite{krim1996two} has been an active research area for decades. This paper focuses on the narrowband far-field source case where the wave front is assumed to be planar and the angle/direction information is to be estimated, known as the direction of arrival (DOA) estimation problem. MUSIC\cite{schmidt1981signal} is the most successful method among conventional DOA estimation techniques. It has been proven to be a realization of the maximum likelihood method in the case of a large number of snapshots and uncorrelated source signals\cite{stoica1989music}. The research on DOA estimation has been advanced in recent years owing to the development of methods based on sparse signal reconstruction (SSR) or compressed sensing (CS) \cite{candes2006compressive}. In these methods, e.g., $\ell_1$-SVD\cite{malioutov2005sparse}, a fixed sampling grid is selected firstly that serves as the set of all candidates of DOA estimates. Then by assuming that all true (unknown) DOAs are exactly on the selected grid, an SSR problem can be formulated where the DOAs of interest constitute the support of the sparse signal to be recovered.

In the case of a single measurement vector (SMV, or a single snapshot), $\ell_1$ optimization \cite{candes2006stable} is a favorable approach to the sparse signal recovery due to its guaranteed recovery accuracy. It has been proven in \cite{candes2006stable} that a sparse signal can be accurately recovered under a so-called restricted isometry property (RIP) condition that requires all columns of the measurement matrix be highly incoherent. In the case of multiple measurement vectors (MMV), the sparse signals at all snapshots share the same support. It has been shown in \cite{eldar2010average} that such joint sparsity can be exploited to improve the averaged recovery success probability under a similar incoherent matrix condition. Sparse Bayesian inference/learning (SBI) \cite{tipping2001sparse, ji2008bayesian, babacan2010bayesian, yang2012bayesian} is another popular method for the sparse signal recovery in CS. In SBI, the signal recovery problem is formulated from a Bayesian perspective while the sparsity information is exploited by assuming a sparse prior for the signal of interest. As an example, a Laplace signal prior leads to a maximum {\em a posteriori} (MAP) optimal estimate that coincides with an optimal solution to the $\ell_1$ optimization\cite{babacan2010bayesian}. In the MMV case, the joint sparsity among different (uncorrelated) snapshots is utilized by assuming the same sparse prior for the signals at all snapshots \cite{wipf2007empirical}. Correlations between snapshots have also been studied in a recent paper \cite{zhang2011sparse}. One merit of SBI is its flexibility in modeling sparse signals that can not only promote the sparsity of its solution, e.g., in \cite{yang2012bayesian}, but also exploit the possible structure of the signal to be recovered, e.g., in \cite{he2009exploiting}. Since the Bayesian inference is a probabilistic method and based on heuristics to some extent, one shortcoming of SBI is that it offers fewer guarantees on the signal recovery accuracy as compared with, for example, $\ell_1$ optimization.

Recent advancements in array signal processing include compressive (CS-) MUSIC \cite{kim2012compressive} and subspace-augmented (SA-) MUSIC \cite{lee2010subspace}. They are combinations of the conventional MUSIC technique and recent CS methods with guaranteed support recovery performance and can outperform MUSIC and standard CS approaches. Though existing CS-based approaches have shown their improvements in DOA estimation, e.g., their success in the case of limited snapshots, there are still difficulties in practical situations where the true DOAs are not on the sampling grid. On one hand, a dense sampling grid is necessary for accurate DOA estimation to reduce the gap between the true DOA and its nearest grid point since the estimated DOAs are constrained on the grid. On the other hand, a dense sampling grid leads to a highly coherent matrix that violates the condition for the sparse signal recovery. We refer to the model adopted in the standard CS methods as an on-grid model hereafter in the sense that the estimated DOAs are constrained on the fixed grid.

An off-grid model for DOA estimation is studied in \cite{zhu2011sparsity} where the estimated DOAs are no longer constrained in the sampling grid set. The model takes into account the basis mismatch in the measurement matrix caused by the off-grid DOAs. It has been shown in \cite{zhu2011sparsity} that the sparse total least squares (STLS) solver proposed in \cite{zhu2011sparsity} can yield an MAP optimal estimate if the matrix perturbation caused by the basis mismatch is Gaussian. However, we show in this paper that the Gaussian condition cannot be satisfied in the off-grid DOA estimation problem and hence a new solver is needed.

In this paper, we propose a Bayesian algorithm for DOA estimation based on the off-grid model that applies to both SMV and MMV cases. The off-grid distance (the distance from the true DOA to the nearest grid point) that lies in a bounded interval is assumed to be uniformly distributed (noninformative) rather than Gaussian as in \cite{zhu2011sparsity}. We refer to our algorithm as off-grid sparse Bayesian inference (OGSBI) in the body of the paper. We then incorporate in our algorithm an idea in \cite{malioutov2005sparse}, using the singular value decomposition (SVD) to reduce the computational workload of the signal recovery process and the sensitivity to noise, namely, OGSBI-SVD. Our approach is a spectral-like method with the value of the spectrum at each DOA being the estimated source power from such a direction. We show by numerical simulations that the proposed method has a smaller mean squared error (MSE) in comparison with $\ell_1$-SVD \cite{malioutov2005sparse}. Indeed, the proposed method can exceed a lower bound of MSE that is shared among all on-grid model based methods including CS-MUSIC and SA-MUSIC. It is also shown that the proposed method is more accurate and faster than STLS. Moreover, the proposed method can estimate DOAs accurately even under a coarse sampling grid.

Notations used in this paper are as follows. Bold-face letters are reserved for vectors and matrices. $\overline{\m{x}}$, $\m{x}^T$ and $\m{x}^H$ denote complex conjugate, transpose and conjugate transpose of a vector $\m{x}$, respectively. $\onen{\cdot}$, $\twon{\cdot}$ and $\frobn{\cdot}$ denote the $\ell_1$ norm, $\ell_2$ norm and Frobenious norm, respectively. $\abs{\m{A}}$, $Tr\lbra{\m{A}}$ are the determinant and trace of a matrix $\m{A}$ respectively. $x_j$ is the $j$th entry of a vector $\m{x}$. $\m{A}_j$, $\m{A}^j$ and $A_{ij}$ are the $j$th column, $j$th row and $(i,j)$th entry of a matrix $\m{A}$, respectively. $\diag\sbra{\m{A}}$ denotes a column vector composed of the diagonal elements of a matrix $\m{A}$, and $\diag\sbra{\m{x}}$ is a diagonal matrix with $\m{x}$ being its diagonal elements. $x'(\theta)$ is the derivative of $x(\theta)$ with respect to $\theta$. $\Re$ and $\Im$ take the real and imaginary parts of a complex variable respectively. $\m{x}\odot\m{y}$ is the Hadamard (element-wise) product of $\m{x}$ and $\m{y}$. $\widehat x$ denotes an estimate of $x$.

The rest of the paper is organized as follows. Section \ref{sec:offgridmodel} studies the off-grid model used in this paper. Section \ref{sec:SBI_MultipleTimeSamples} introduces the proposed OGSBI and OGSBI-SVD algorithms. Section \ref{sec:simulation} presents our simulation results. Section \ref{sec:discussion} concludes this paper.

\section{Off-grid DOA Estimation Model} \label{sec:offgridmodel}

Consider $K$ narrowband far-field sources $s_k(t)$, $k=1,\cdots,K$, impinging on an array of $M$ omnidirectional sensors from directions $\theta_k$, $k=1,\cdots,K$. Time delays at different sensors can be represented by simple phase shifts, leading to the observation model\cite{krim1996two}:
\equ{\m{y}(t) = \m{A}(\m{\theta})\m{s}(t)+\m{e}(t),\quad t=1, \cdots, T,\label{formu:observation_model}}
where $\m{y}(t)=\mbra{y_1(t),\cdots,y_M(t)}^T$, $\m{\theta}=\mbra{\theta_1,\cdots,\theta_K}^T$, $\m{s}(t)=\mbra{s_1(t),\cdots,s_K(t)}^T$, $\m{e}(t)=\mbra{e_1(t),\cdots,e_M(t)}^T$, and $y_m(t)$ and $e_m(t)$, $m = 1,\cdots,M$, are the output and measurement noise of the $m$th sensor at time $t$ respectively. The matrix $\m{A}(\m{\theta})=\mbra{\m{a}\sbra{\theta_1},\cdots,\m{a}\sbra{\theta_K}}$ is an array manifold matrix and $\m{a}\sbra{\theta_k}$ is called steering vector of the $k$th source. The entry $\m{a}_m\sbra{\theta_k}$ contains the delay information of the $k$th source to the $m$th sensor. In this paper, we assume that the number of sources $K$ is already known. Readers are referred to a preprint version \cite{yang2011off} for discussions on the case of unknown $K$. So, the goal is to find the unknown DOAs $\m{\theta}$ given $K$, $\m{y}(t)$ and the mapping $\m{\theta}\rightarrow \m{A}(\m{\theta})$. In the following we re-derive the off-grid model proposed in \cite{zhu2011sparsity} using linear approximation and further show its relationship with the on-grid one.

Let $\tilde{\m{\theta}}=\lbra{\tilde{\theta}_1, \cdots, \tilde{\theta}_N}$ be a fixed sampling grid in the DOA range $\mbra{0,\pi}$, where $N$ denotes the grid number and typically satisfies $N\gg M > K$. Without loss of generality, let $\tilde{\m{\theta}}$ be a uniform grid with a grid interval $r=\tilde{\theta}_2-\tilde{\theta}_1\propto N^{-1}$.
Suppose $\theta_k\notin\lbra{\tilde{\theta}_1, \cdots, \tilde{\theta}_N}$ for some $k\in\lbra{1,\cdots,K}$ and that $\tilde{\theta}_{n_k}$, $n_k\in\lbra{1, \cdots, N}$, is the nearest grid point to $\theta_k$. We approximate the steering vector $\m{a}\sbra{\theta_k}$ using linearization:
\equ{\m{a}\sbra{\theta_k}\approx\m{a}\sbra{\tilde{\theta}_{n_k}}+\m{b}\sbra{\tilde{\theta}_{n_k}} \sbra{\theta_{k} -\tilde{\theta}_{n_k}} \label{formu:offgridlinearization}}
with $\m{b}\sbra{\tilde{\theta}_{n_k}}=\m{a}'\sbra{\tilde{\theta}_{n_k}}$. Denote $\m{A}=\mbra{\m{a}\sbra{\tilde{\theta}_1},\cdots,\m{a}\sbra{\tilde{\theta}_N}}$, $\m{B}=\mbra{\m{b}\sbra{\tilde{\theta}_1},\cdots,\m{b}\sbra{\tilde{\theta}_N}}$, $\m{\beta}=\mbra{\beta_1,\cdots,\beta_N}^T\in\mbra{-\frac{1}{2}r,\frac{1}{2}r}^N$ and $\m{\Phi}\sbra{\m{\beta}}=\m{A}+\m{B}\diag\sbra{\m{\beta}}$, where for $n=1,\cdots,N$,
\equ{\begin{split}
&\beta_n=\theta_k-\tilde{\theta}_{n_k}, \\
&x_n(t)=s_{n_k}(t), \quad\text{if }n=n_k \text{ for any } k\in\lbra{1,\cdots,K};\\
&\beta_n=0,\quad x_n(t)=0,\quad\text{otherwise},\end{split}}
with $n_k\in\lbra{1,\cdots,N}$ and $\tilde{\theta}_{n_k}$ being the nearest grid to a source $\theta_k$, $k\in\lbra{1,\cdots,K}$. By absorbing the approximation error into the measurement noise the observation model in (\ref{formu:observation_model}) can be written into
\equ{\m{y}(t) = \m{\Phi}\sbra{\m{\beta}}\m{x}(t)+\m{e}(t),\quad t=1, \cdots, T,\label{formu:offgridmodel}}
which is the off-grid model to be used in this paper. This model will be empirically validated in Subsection \ref{sec:ComparisonWithL1SVD} by showing that the total noise (approximation error plus measurement noise) follows the Gaussian distribution with high probability if the measurement noise is Gaussian.

It should be noted that the off-grid model in (\ref{formu:offgridmodel}) is closely related to the on-grid one that can be obtained by setting $\m{\beta}=\m{0}$ in (\ref{formu:offgridmodel}) ($\m{\Phi}\sbra{\m{0}}=\m{A}$). In fact, the off-grid model can be considered as the first order approximation of the true observation model while the on-grid one is the zeroth order approximation. As a result, the off-grid model has a much smaller modeling error than the on-grid one. Such an advantage is twofold. First, by adopting the same sampling grid the off-grid model results in higher accuracy, especially in the case of a low measurement noise where the modeling error is the dominant modeling uncertainty. Second, a coarser sampling grid can be adopted in the off-grid model to achieve a considerably reduced computational workload with a comparable modeling accuracy.

To estimate the DOAs $\m{\theta}$ we need to find not only the support of the sparse signals $\m{x}(t)$, $t=1,\cdots,T$, but also the off-grid difference $\m{\beta}$. In this paper, we formulate the problem based on a Bayesian perspective and develop an iterative algorithm to jointly estimate $\m{x}(t)$, $t=1,\cdots,T$, and $\m{\beta}$ in the following section.

\section{OGSBI: Off-grid Sparse Bayesian Inference} \label{sec:SBI_MultipleTimeSamples}
We consider complex-valued signals throughout the paper since the matrix $\m{\Phi}\sbra{\m{\beta}}$ is complex-valued. We derive our algorithm in the MMV case. The SMV is a special case by simply setting $T=1$. Denote $\m{Y}=\mbra{\m{y}(1),\cdots,\m{y}(T)}$, $\m{X}=\mbra{\m{x}(1),\cdots,\m{x}(T)}$ and $\m{E}=\mbra{\m{e}(1),\cdots,\m{e}(T)}$. The off-grid DOA estimation model in (\ref{formu:offgridmodel}) becomes
\equ{\m{Y} = \m{\Phi}\sbra{\m{\beta}}\m{X}+\m{E}}
with $\m{\Phi}\sbra{\m{\beta}}=\m{A}+\m{B}\diag\sbra{\m{\beta}}$, $\m{Y},\m{E}\in\bC^{M\times T}$, $\m{X}\in\bC^{N\times T}$, $\m{A},\m{B},\m{\Phi}\sbra{\m{\beta}}\in\bC^{M\times N}$ and $\m{\beta}\in\mbra{-\frac{1}{2}r,\frac{1}{2}r}^N$. The matrix $\m{X}$ of interest is jointly sparse (or row-sparse), i.e., all columns of $\m{X}$ are sparse and share the same support.

\subsection{Sparse Bayesian Formulation}

\subsubsection{Noise model}
Under an assumption of white (circular symmetric) complex Gaussian \cite{goodman1963statistical} noises, we have
\equ{p\sbra{\m{E}|\alpha_0}=\prod_{t=1}^T\ccn\sbra{\m{e}(t)|\m{0},\alpha_0^{-1}\m{I}}\label{formu:prior_noise}}
where $\alpha_0=\sigma^{-2}$ denotes the noise precision with $\sigma^2$ being the noise variance, the probability density function (PDF) of a (circular symmetric) complex Gaussian distributed random variable $\m{u}\sim\ccn\sbra{\m{\mu},\m{\Sigma}}$ with mean $\m{\mu}$ and covariance $\m{\Sigma}$ is\cite{goodman1963statistical}
\equ{\ccn\sbra{\m{u}|\m{\mu},\m{\Sigma}}=\frac{1}{\pi^N\abs{\m{\Sigma}}} \exp\lbra{-\sbra{\m{u}-\m{\mu}}^H\m{\Sigma}^{-1}\sbra{\m{u}-\m{\mu}}}.}
Then we have
\equ{p\sbra{\m{Y}|\m{X},\alpha_0,\m{\beta}}=\prod_{t=1}^T \ccn\sbra{\m{y}(t)|\m{\Phi}\sbra{\m{\beta}}\m{x}(t),\alpha_0^{-1}\m{I}}. \label{formu:prior_Y}}


In this paper we assume that the noise precision $\alpha_0$ is unknown. A Gamma hyperprior is assumed for $\alpha_0$ since it is a conjugate prior of the Gaussian distribution:
\equ{p\sbra{\alpha_0;c,d}=\Gamma\sbra{\alpha_0|c,d}\label{formu:prior_alpha0}}
where $\Gamma\sbra{\alpha_0|c,d}=\mbra{\Gamma\sbra{c}}^{-1}d^c{\alpha_0}^{c-1}\exp\lbra{-d\alpha_0}$ with $\Gamma\sbra{\cdot}$ being the Gamma function. We set $c,d\rightarrow0$ as in\cite{tipping2001sparse,ji2008bayesian} to obtain a broad hyperprior.

\subsubsection{Sparse signal model}
A sparse prior is needed for the jointly sparse matrix $\m{X}$ of interest. We assume that the signals among snapshots are independent and adopt the two-stage hierarchical prior:
$p\sbra{\m{X};\rho}=\int p\sbra{\m{X}|\m{\alpha}} p\sbra{\m{\alpha};\rho}\,d\m{\alpha}$,
where $\rho>0$, $\m{\alpha}\in\bR^N$, $\m{\Lambda}=\diag\sbra{\m{\alpha}}$ and
{\lentwo\equa{p\sbra{\m{X}|\m{\alpha}}
&=& \prod_{t=1}^T\ccn\sbra{\m{x}(t)|\m{0},\m{\Lambda}},\label{formu:prior_X}\\ p\sbra{\m{\alpha};\rho}
&=& \prod_{n=1}^N\Gamma\sbra{\alpha_n|1,\rho}.\label{formu:prior_alpha}
}}It is easy to show that all columns of $\m{X}$ are independent and share the same prior. According to \cite{babacan2010bayesian}, for $t=1,\cdots,T$ both $\Re\lbra{\m{x}(t)}$ and $\Im\lbra{\m{x}(t)}$ are Laplace distributed and share the same PDF that is strongly peaked at the origin. As a result, the two-stage hierarchical prior is a sparse prior that favors most rows of $\m{X}$ being zeros.


\subsubsection{Off-grid distance model}
We assume a uniform prior for $\m{\beta}$:
\equ{\m{\beta}\sim U\sbra{\mbra{-\frac{1}{2}r,\frac{1}{2}r}^N}. \label{formu:prior_beta}}
The prior is noninformative in the sense that the only information of $\m{\beta}$ we use is its boundedness.

By combining the stages of the hierarchical Bayesian model, the joint PDF is
\equ{p\sbra{\m{X},\m{Y},\alpha_0,\m{\alpha},\m{\beta}}= p\sbra{\m{Y}|\m{X},\alpha_0,\m{\beta}}p\sbra{\m{X}|\m{\alpha}}p\sbra{\m{\alpha}}p\sbra{\alpha_0}p\sbra{\m{\beta}} \label{formu:jointdistributionforall}}
with the distributions on the right hand side as defined by (\ref{formu:prior_Y}), (\ref{formu:prior_X}), (\ref{formu:prior_alpha}), (\ref{formu:prior_alpha0}) and (\ref{formu:prior_beta}) respectively.

\subsection{Bayesian Inference}

An evidence procedure \cite{mackay1992bayesian} is exploited to perform the Bayesian inference since the exact posterior distribution $p\sbra{\m{X},\alpha_0,\m{\alpha},\m{\beta}|\m{Y}}$ cannot be explicitly calculated. Similar approaches have been used in standard Bayesian CS methods \cite{ji2008bayesian,babacan2010bayesian}. First it is easy to show that the posterior distribution of $\m{X}$ is a complex Gaussian distribution:
\equ{p\sbra{\m{X}|\m{Y},\alpha_0,\m{\alpha},\m{\beta}}= \prod_{t=1}^T\ccn\sbra{\m{x}(t)|\m{\mu}(t),\m{\Sigma}}\label{formu:posterior_X}}
with
\lentwo{\equa{\m{\mu}(t)
&=&\alpha_0\m{\Sigma}\m{\Phi}^H\m{y}(t),\quad t=1,\cdots,T, \label{formu:mu_postX}\\\m{\Sigma}
&=&\sbra{\alpha_0\m{\Phi}^H\m{\Phi}+\m{\Lambda}^{-1}}^{-1}.\label{formu:Sigma_postX}}
}Calculations of $\m{\Sigma}$ and $\m{\mu}(t)$, $t=1,\cdots,T$, need estimates of the hyperparameters $\alpha_0$, $\m{\alpha}$ and $\m{\beta}$. In an evidence procedure, they are estimated using an MAP estimate that maximizes $p\sbra{\alpha_0,\m{\alpha},\m{\beta}|\m{Y}}$. It can be easily observed that to maximize $p\sbra{\alpha_0,\m{\alpha},\m{\beta}|\m{Y}}$ is equivalent to maximizing the joint PDF $p\sbra{\m{Y},\alpha_0,\m{\alpha},\m{\beta}}= p\sbra{\alpha_0,\m{\alpha},\m{\beta}|\m{Y}}p\sbra{\m{Y}}$ since $p\sbra{\m{Y}}$ is independent of the hyperparameters.
An expectation-maximization (EM) algorithm is implemented that treats $\m{X}$ as a hidden variable and turns to maximizing  $E\lbra{\log p\sbra{\m{X},\m{Y},\alpha_0,\m{\alpha},\m{\beta}}}$, where $p\sbra{\m{X},\m{Y},\alpha_0,\m{\alpha},\m{\beta}}$ is given in (\ref{formu:jointdistributionforall}) and $E\lbra{\cdot}$ denotes an expectation with respect to the posterior of $\m{X}$ as given in (\ref{formu:posterior_X}) using the current estimates of the hyperparameters.

Denote $\m{\cU}=\mbra{\m{\mu}(1),\cdots,\m{\mu}(T)}=\alpha_0\m{\Sigma}\m{\Phi}^H\m{Y}$, $\underline{\m{X}}=\m{X}/\sqrt{T}$, $\underline{\m{Y}}=\m{Y}/\sqrt{T}$, $\underline{\m{\cU}}=\m{\cU}/\sqrt{T}$ and $\underline{\rho}=\rho/T$. Following a similar procedure as in \cite{tipping2001sparse}, it is easy to obtain the following updates of $\m{\alpha}$ and $\alpha_0$:
{\lentwo\equa{\alpha_n^{new}
&=&\frac{\sqrt{1+4\underline{\rho} E\lbra{\twon{\underline{\m{X}}^n}^2}}-1}{2\underline{\rho}}, \quad n=1,\cdots,N,\label{formu:update_alpha_Multiple}\\
\alpha_0^{new}
&=&\frac{M+(c-1)/T}{ E\lbra{\frobn{\underline{\m{Y}}-\m{\Phi}\underline{\m{X}}}^2}+d/T},\label{formu:update_alpha0_Multiple}}
}where $ E\lbra{\twon{\underline{\m{X}}^n}^2}=\twon{\underline{\m{\cU}}^n}^2+\Sigma_{nn}$, $ E\lbra{\frobn{\underline{\m{Y}}-\m{\Phi}\underline{\m{X}}}^2}=\frobn{\underline{\m{Y}}-\m{\Phi}\underline{\m{\cU}}}^2+\alpha_0^{-1} \sum_{n=1}^N\gamma_n$ with $\gamma_n=1-\alpha_n^{-1}\Sigma_{nn}$.

For $\m{\beta}$, its estimate maximizes $E\lbra{\log p\sbra{\m{Y}|\m{X},\alpha_0,\m{\beta}}p\sbra{\m{\beta}}}$ by (\ref{formu:jointdistributionforall}) and thus minimizes
\equ{\begin{split}
&E\lbra{\frac{1}{T}\sum_{t=1}^T\twon{\m{y}(t)-\sbra{\m{A}+\m{B}\diag\sbra{\m{\beta}}}\m{x}(t)}^2}\\
&=\frac{1}{T}\sum_{t=1}^T\twon{\m{y}(t)-\sbra{\m{A}+\m{B}\diag\sbra{\m{\beta}}}\m{\mu}(t)}^2\\
&\quad+Tr\lbra{\sbra{\m{A}+\m{B}\diag\sbra{\m{\beta}}}\m{\Sigma}\sbra{\m{A}+\m{B}\diag\sbra{\m{\beta}}}^H}\\
&=\m{\beta}^T \m{P}\m{\beta}-2\m{v}^T\m{\beta}+C
\end{split}\label{formu:solve_beta}}
where $C$ is a constant term independent of $\m{\beta}$, $\m{P}$ is a positive semi-definite matrix and
{\lentwo\equa{\m{P}
&=&\Re\lbra{\overline{\m{B}^H\m{B}}\odot\sbra{\underline{\m{\cU}}\cdot\underline{\m{\cU}}^H +\m{\Sigma}}},\\\m{v}
&=&\Re\lbra{\frac{1}{T}\sum_{t=1}^T\diag\sbra{\overline{\m{\mu}(t)}}\m{B}^H\sbra{\m{y}(t) -\m{A}\m{\mu}(t)}}\notag\\
&\quad&-\Re\lbra{\diag\sbra{\m{B}^H\m{A}\m{\Sigma}}}.}
}The detailed derivation of (\ref{formu:solve_beta}) is provided in Appendix for simplicity of exposition. As a result, we have
\equ{\m{\beta}^{new}=\arg\min_{\m{\beta}\in\mbra{-\frac{1}{2}r,\frac{1}{2}r}^N}\lbra{\m{\beta}^T \m{P}\m{\beta}-2\m{v}^T\m{\beta}}.\label{formu:update_beta}}
\begin{rem}
Though an explicit expression of $\m{\beta}^{new}$ cannot be given, by recognizing that $\m{\beta}$ is jointly sparse with $\m{x}$, the dimension of $\m{\beta}$ can be reduced to $K$ in the computation and hence $\m{\beta}^{new}$ can be efficiently calculated. We provide the details in Subsection \ref{sec:implementation}.
\end{rem}

The proposed OGSBI algorithm is implemented as follows. After initializations of the hyperparameters $\m{\alpha}$, $\alpha_0$ and $\m{\beta}$, we calculate $\m{\Sigma}$ and $\m{\mu}(t)$, $t=1,\cdots,T$, using the current values of the hyperparameters according to (\ref{formu:Sigma_postX}) and (\ref{formu:mu_postX}) respectively. Then we update $\m{\alpha}$, $\alpha_0$ and $\m{\beta}$ according to (\ref{formu:update_alpha_Multiple}), (\ref{formu:update_alpha0_Multiple}) and (\ref{formu:update_beta}) respectively. The process is repeated until some convergence criterion is satisfied. We note that OGSBI has guaranteed convergence since the function $p\sbra{\alpha_0,\m{\alpha},\m{\beta}|\m{Y}}$ is guaranteed to increase at each iteration by the property of EM algorithm\cite{mclachlan1997algorithm}.

\subsection{OGSBI-SVD}\label{sec:SBI_Multiple:SBISVD}
In this subsection we recall a subspace-based idea in \cite{malioutov2005sparse} that uses the SVD of the measurement matrix $\m{Y}=\m{U}\m{S}\m{V}^H$ to reduce the computation of the signal reconstruction process and the sensitivity to the measurement noise. Then we incorporate it into our OGSBI algorithm.
Consider the noise-free case where $\m{Y}=\m{\Phi}\m{X}$ with $K\leq T$. We have $\text{Rank}\sbra{\m{Y}}\leq\text{Rank}\sbra{\m{X}}\leq K$. Let $\m{V}=\mbra{\m{V}_1\text{ }\m{V}_2}$, where $\m{V}_1$ and $\m{V}_2$ are matrices that consist of the first $K$ and the rest $T-K$ columns of $\m{V}$ respectively. Then we have that $\m{Y}_{SV}=\m{Y}\m{V}_1\in\bC^{M\times K}$ preserves all signal information. In a general case where noises exist, by the SVD we have $\m{Y}\m{V}=\mbra{\m{Y}_{SV}\,\,\m{Y}\m{V}_2}$, where the first part $\m{Y}_{SV}$ preserves most signal information and is to be used in the following signal recovery process while the second part is abandoned. Denote $\m{X}_{SV}=\m{X}\m{V}_1$ and $\m{E}_{SV}=\m{E}\m{V}_1$. Then we have
\equ{\m{Y}_{SV}=\m{\Phi}\m{X}_{SV}+\m{E}_{SV}.\label{formu:offgridmodel_SVD}}
In (\ref{formu:offgridmodel_SVD}), $\m{Y}_{SV}$, $\m{X}_{SV}$ and $\m{E}_{SV}$ can be viewed as the new matrices of sensor measurements, source signals and measurement noises respectively. The joint sparsity still holds in $\m{X}_{SV}$. We do not exploit possible correlations that exist between columns of $\m{X}_{SV}$ (and in $\m{E}_{SV}$), i.e., we still assume that $\m{X}_{SV}$ (and $\m{E}_{SV}$) have independent columns.\footnote{The correlations between columns of the signal matrix ($\m{X}_{SV}$ in our case) have recently been studied in \cite{zhang2011sparse}.} It is then straightforward to apply the proposed OGSBI algorithm to estimate $\m{X}_{SV}$, $\m{\beta}$ and then the DOAs. We use OGSBI-SVD to refer to the resulting algorithm.

Based on implementation details to be introduced in Subsection \ref{sec:implementation}, it can be shown that OGSBI-SVD has a computational complexity of order $O\sbra{MN^2}$ per iteration while that for OGSBI is $O\sbra{\max\sbra{MN^2,MNT}}$ per iteration. An additional computational workload of order $O\sbra{\max\sbra{M^2T,MT^2}}$ is for the SVD of $\m{Y}$ in OGSBI-SVD. Since it is empirically found that OGSBI-SVD converges much faster than OGSBI, the whole computational workload of OGSBI-SVD is less than that of OGSBI in general.\footnote{A possible exception happens in the case of $T\gg N$ where the computation for the SVD is quite heavy. A modified approach in such a case is to partition $\m{Y}$ firstly into blocks with each of about $N$ columns, then operate the SVD on each block and keep the resulting signal subspaces, and finally do another SVD on the new signal matrix composed of all signal subspaces. A model similar to (\ref{formu:offgridmodel_SVD}) can be cast.}

\subsection{Source Power and DOA Estimation}
We use the estimated source powers from different directions to form a spectrum of the proposed algorithm. In the following we derive a formula to estimate the source powers. We take OGSBI-SVD as an example. The case of OGSBI is similar with some modifications \cite{yang2011off}. Let $\widehat{\m{X}}=\m{X}_{SV}\m{V}_1^H$ be an estimate of the signal $\m{X}$. Then consider $\widehat{\m{X}}$ row by row and we have $\widehat{\m{X}}^n\sim\ccn\sbra{\widehat{\m{\cU}}^n\m{V}_1^H, \widehat{\Sigma}_{nn}\m{V}_1\m{V}_1^H}$ where we use $\widehat{\m{\cU}}$ and $\widehat{\m{\Sigma}}$ to denote the final estimates of the mean and covariance of $\m{X}_{SV}$ respectively. We use the expectation as an estimate of the power $\wp_n$ from direction $\tilde{\theta}_n$ (with a modification of $\beta_n$):
\equ{\begin{split}\widehat{\wp}_n
&=E\lbra{\wp_n}=\frac{1}{T}E\lbra{\twon{\widehat{\m{X}}^n}^2} \\
&=\frac{1}{T}\sbra{\twon{E\lbra{\widehat{\m{X}}^n}}^2+E\lbra{\twon{\widehat{\m{X}}^n-E\lbra{\widehat{\m{X}}^n}}^2}}\\
&=\frac{1}{T}\sbra{\twon{\widehat{\m{\cU}}^n\m{V}_1^H}^2+Tr\lbra{\widehat{\Sigma}_{nn}\m{V}_1\m{V}_1^H}}\\
&=\frac{\twon{\widehat{\m{\cU}}^n}^2}{T}+\frac{K\widehat{\Sigma}_{nn}}{T}.\end{split}}

Like other spectral-based methods, the DOAs are estimated using the locations of the highest peaks of the spectrum. Suppose that the grid indices of the highest $K$ peaks of $\widehat{\m{\wp}}$ are $\widehat{n}_k$, $k=1,\cdots,K$. The estimated $K$ DOAs will be $\widehat{\theta}_k=\tilde{\theta}_{\widehat n_k}+\widehat{\beta}_{\widehat n_k}$, $k=1,\cdots,K$.

\subsection{Implementation Details}\label{sec:implementation}
This subsection presents some details of our implementations of OGSBI and OGSBI-SVD. At each iteration of OGSBI or OGSBI-SVD, an $N\times N$ matrix inversion is required when updating $\m{\Sigma}$ according to (\ref{formu:Sigma_postX}). By $M<N$ the Woodbury matrix identity is applied to give
$\m{\Sigma}=\m{\Lambda}-\m{\Lambda}\m{\Phi}^H\m{C}^{-1}\m{\Phi}\m{\Lambda}$
with $\m{C}=\alpha_0^{-1}\m{I}+\m{\Phi}\m{\Lambda}\m{\Phi}^H\in\bC^{M\times M}$.

By the fact that $\m{\beta}$ is jointly sparse with $\m{x}(t)$ whose $K$ nonzero entries correspond to the locations of the $K$ sources, we calculate only entries of $\m{\beta}$ that correspond to locations of the maximum $K$ entries of $\m{\alpha}$ and set others to zeros. As a result, $\m{\beta}$, $\m{P}$ and $\m{v}$ can be truncated into dimension of $K$ or $K\times K$. We still use $\m{\beta}$, $\m{P}$ and $\m{v}$ hereafter to denote their truncated versions for simplicity. By (\ref{formu:update_beta}) and  $\frac{\partial}{\partial\m{\beta}}\lbra{\m{\beta}^T \m{P}\m{\beta}-2\m{v}^T\m{\beta}}=2\sbra{\m{P}\m{\beta}-\m{v}}$ we have $\m{\beta}^{new}=\check{\m{\beta}}$ if $\m{P}$ is invertible and $\check{\m{\beta}}=\m{P}^{-1}\m{v}\in\mbra{-\frac{1}{2}r,\frac{1}{2}r}^K$. Otherwise, we update $\m{\beta}$ elementwise, i.e., at each step we update one $\beta_n$ by fixing up the other entries of $\m{\beta}$. For $n=1,\cdots,K$, first we let
\equ{\check{\beta}_n=\frac{v_n-\sbra{\m{P}_n}_{-n}^T\m{\beta}_{-n}}{P_{nn}},\label{formu:update_betahat}}
where $\m{u}_{-n}$ is $\m{u}$ without the $n$th entry for a vector $\m{u}$. Then by constraining $\beta_n\in\mbra{-\frac{1}{2}r, \frac{1}{2}r}$ we have
\equ{\beta_n^{new}=\left\{\begin{array}{ll}
   \check{\beta}_n, & \text{ if }\check{\beta}_n\in\mbra{-\frac{1}{2}r,\frac{1}{2}r};\\
   -\frac{1}{2}r, & \text{ if }\check{\beta}_n<-\frac{1}{2}r;\\
   \frac{1}{2}r, & \text{ otherwise}.\end{array}\right. \label{formu:betan}}
It is easy to show that the objective function is guaranteed to decrease at each step with $\beta_n$ defined in (\ref{formu:betan}).

We terminate OGSBI and OGSBI-SVD if $\frac{\twon{\m{\alpha}^{i+1}-\m{\alpha}^i}}{\twon{\m{\alpha}^i}}<\tau$ or the maximum number of iterations is reached, where $\tau$ is a user-defined tolerance and the superscript $i$ refers to the iteration.


\section{Numerical simulations}\label{sec:simulation}
In this section, we present our numerical results for the DOA estimation. A standard uniform linear array (ULA) of $M=8$ sensors is considered. The origin is set at the middle point of the ULA to reduce the approximation error in (\ref{formu:offgridlinearization}). So we have $A_{mn}=\exp\lbra{j\pi\sbra{m-\frac{M+1}{2}} \cos\tilde{\theta}_n}$ and $B_{mn}=-j\pi\sbra{m-\frac{M+1}{2}}\sin\tilde{\theta}_n\cdot A_{mn}$, $m=1,\cdots,M$, $n=1,\cdots,N$, with $j=\sqrt{-1}$. A uniform sampling grid $\lbra{0^\circ,r,2r,\cdots,180^\circ}$ is considered with $r$ being the grid interval. The number of snapshots is set to $T=200$ in the case of MMV. We consider only OGSBI-SVD in the MMV case since it is empirically observed to converge faster and be more accurate in comparison with OGSBI. In OGSBI-SVD, we set $\rho=0.01$ and $c=d=1\times10^{-4}$. We initialize $\alpha_0=\frac{100K}{\sum_{t=1}^K Var\lbra{\sbra{\m{Y}_{SV}}_t}}$, $\m{\alpha}=\frac{1}{MK}\sum_{t=1}^K\abs{\m{A}^H\sbra{\m{Y}_{SV}}_t}$ and $\m{\beta}=\m{0}$, where $\abs{\cdot}$ applies elementwise. We set $\tau=10^{-3}$ and the maximum number of iterations to 1000. We note that the proposed algorithm is insensitive to the initializations of $\alpha_0$, $\m{\alpha}$ and $\m{\beta}$, as well as to $\rho$ if $\rho$ is not too large. As reported in \cite{wipf2007empirical}, the estimate of $\alpha_0$ can be inaccurate in some cases. But we have observed minimal effects on the result of DOA estimation. Readers are referred to \cite{yang2011off} for detailed discussions. All experiments are carried out in Matlab v.7.7.0 on a PC with a Windows XP system and a 3GHz CPU. Matlab codes have been made available online at \href{https://sites.google.com/site/zaiyang0248/publication}
{https://sites.google.com/site/zaiyang0248/publication}.

\subsection{Comparison with $\ell_1$-SVD} \label{sec:ComparisonWithL1SVD}
\begin{figure}
\centering
  \includegraphics[width=3.5in]{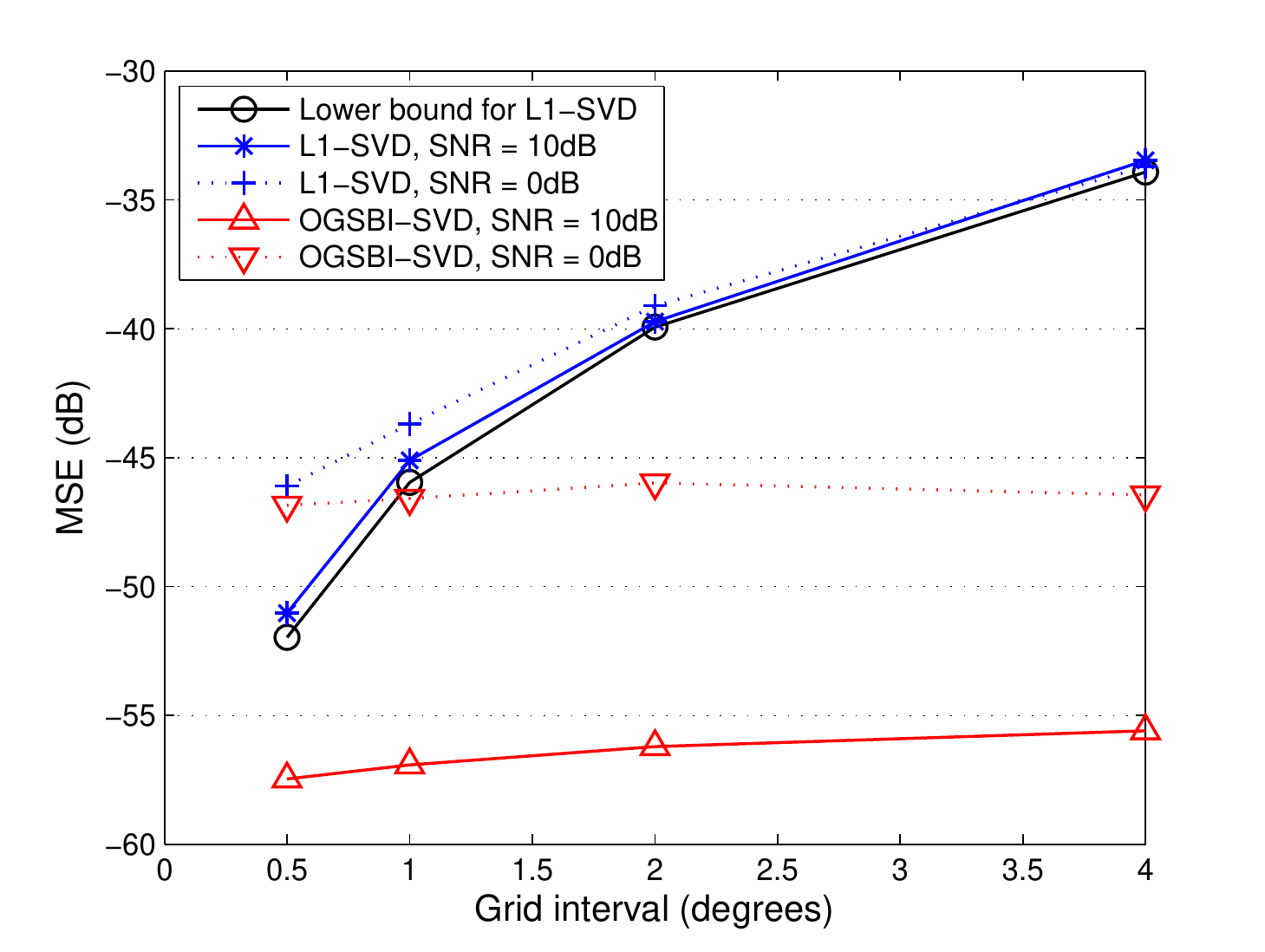}
\centering
\caption{MSEs of OGSBI-SVD and $\ell_1$-SVD. The lower bound is for $\ell_1$-SVD regardless of the SNR.} \label{Fig:ComparisonMSE}
\end{figure}

We take $\ell_1$-SVD in \cite{malioutov2005sparse} as a representative of on-grid model based methods and compare OGSBI-SVD with it in terms of mean squared error (MSE) and computational time with respect to the grid interval $r$ and SNR. In our experiment, we consider $\text{SNR}=10$ and $0$dB, and $r=0.5^\circ,1^\circ,2^\circ\text{ and }4^\circ$. In each trial, $K=2$ sources $\theta_1$, $\theta_2$ are uniformly generated within intervals $\mbra{58^\circ,62^\circ}$ and $\mbra{86^\circ,90^\circ}$ respectively. Before presenting our comparison results, we show using Kolmogorov-Smirnov test that the total noise (measurement noise plus approximation error) in (\ref{formu:offgridmodel}) is Gaussian distributed with a rate of at least $94\%$ in all scenarios. This empirically validates the off-grid model. For each combination $\sbra{\text{SNR},r}$, the MSE is averaged over $R=200$ trials:
$\text{MSE}=\frac{1}{RK}\sum_{i=1}^R\sum_{k=1}^K\sbra{\theta_k^i-\widehat{\theta}_k^i}^2 $
where the superscript $i$ refers to the $i$th trial. It should be noted that there exists a lower bound for the MSE of $\ell_1$-SVD regardless of the SNR since the best DOA estimate that $\ell_1$-SVD can obtain is the grid point nearest to the true DOA. In fact, the lower bound is shared among all on-grid model based methods including CS-MUSIC, SA-MUSIC and the algorithm in \cite{zhang2011sparse}. By assuming that the true DOA is uniformly distributed, the lower bound can be easily calculated as
$\text{LB}=r^2/12$.\footnote{The presented lower bound is, in fact, the expectation in the case of limited trials. The variance approaches zero as the number of trials gets large.}
Fig. \ref{Fig:ComparisonMSE} presents our experimental results. In all scenarios under consideration, OGSBI-SVD has more accurate DOA estimation than $\ell_1$-SVD. Moreover, OGSBI-SVD can exceed the lower bound for $\ell_1$-SVD in most scenarios. The phenomenon is significant in the case of a higher SNR or a coarser sampling grid where the on-grid model has a poor performance on describing the true observation model while it is overcome to a large extent by the off-grid model used in this paper.

Table \ref{table:Time} presents the averaged CPU times of OGSBI-SVD and $\ell_1$-SVD (excluding the SVD process that takes about $0.003$s in our case) with respect to SNR and $r$.\footnote{The code of $\ell_1$-SVD is provided by the author of \cite{malioutov2005sparse}. We note that its speed can be accelerated using state-of-the-art algorithms for CS.} For both OGSBI-SVD and $\ell_1$-SVD, their CPU times decrease as the grid gets coarser. OGSBI-SVD is faster than $\ell_1$-SVD at $r=2^\circ$ and $4^\circ$. One drawback of the proposed method is that it is slow in the case of a dense sampling grid. In practice, we recommend to use a coarser grid with $r=2^\circ$ for the proposed algorithm since it can give an accurate yet fast DOA estimation.

\begin{table}
 \caption{Averaged Time Consumptions of $\ell_1$-SVD and OGSBI-SVD With Respect to SNR and $r$. Time unit: sec.}
 \centering
\begin{tabular}{l|r|r|r|r}
  \hline\hline
  \multicolumn{5}{c}{$\text{SNR}=10$dB}\\
  \hline
   & $r=0.5^\circ$ & $r=1^\circ$ & $r=2^\circ$ & $r=4^\circ$ \\\hline
  $\ell_1$-SVD & $0.601$ & $0.413$ & $0.324$ & $0.291$ \\
  OGSBI-SVD & $10.2$ & $0.782$ & $0.096$ & $0.025$ \\
  \hline\hline
  \multicolumn{5}{c}{$\text{SNR}=0$dB}\\
  \hline
   & $r=0.5^\circ$ & $r=1^\circ$ & $r=2^\circ$ & $r=4^\circ$ \\\hline
  $\ell_1$-SVD & $0.413$ & $0.295$ & $0.218$ & $0.190$ \\
  OGSBI-SVD & $10.9$ & $0.831$ & $0.104$ & $0.024$ \\
  \hline\hline
\end{tabular}\label{table:Time}
\end{table}


\begin{rem}~
 \begin{itemize}
  \item[(1)] We choose $\ell_1$ optimization for comparison because it is typically known to have better theoretical guarantee for sparse recovery, though simpler solvers in CS, e.g., OMP \cite{tropp2007signal}, may succeed in our setting where the two sources are well separated and have lower computational cost in such low dimensional problems. Readers are referred to \cite{yang2011off} for more simulation results of $\ell_1$ optimization and the proposed method in the case of closely spaced sources.
  \item[(2)] Another advantage of the proposed algorithm is its smaller DOA estimation bias in comparison with that of $\ell_1$-SVD\cite{yang2011off}.
 \end{itemize}
\end{rem}

\subsection{Comparison with STLS}
The off-grid model has recently been used in \cite{zhu2011sparsity} for DOA estimation. In \cite{zhu2011sparsity}, a sparse total least-squares (STLS) approach is proposed. In the SMV case, STLS seeks to solve the nonconvex optimization problem
\equ{\min_{\m{x},\m{\beta}}\lbra{\twon{\m{\beta}}^2+\twon{\m{y}-\sbra{\m{A}+\m{B}\diag\sbra{\m{\beta}}}\m{x}}^2 +\lambda\onen{\m{x}}} \label{formu:problem_STLS}}
where $\m{x}$ is the sparse source signal of interest, $\m{y}$ is the noisy measurement, $\m{A}$, $\m{B}$ and $\m{\beta}$ are the same as defined in the off-grid model, and $\lambda>0$ is a regularization parameter. From the Bayesian perspective, this is equivalent to seeking for an MAP solution of $\sbra{\m{x},\m{\beta}}$ by assuming that the measurement noise is white Gaussian, $\m{x}$ is Laplacian and $\m{\beta}$ is Gaussian. It is noted that the last assumption for $\m{\beta}$ cannot properly capture the property of $\m{\beta}$. A local minima of the problem in (\ref{formu:problem_STLS}) is achieved in \cite{zhu2011sparsity} by an alternating approach, i.e., alternatively solving $\m{x}$ with a fixed $\m{\beta}$, which requires a solution to an $\ell_1$-regularized least square problem, and solving $\m{\beta}$ with a fixed $\m{x}$, which requires a solution to an $N$ dimensional linear system. As argued in \cite{malioutov2005sparse} the SVD used in OGSBI-SVD can alleviate the sensitivity to the measurement noise in the MMV case that is not used in \cite{zhu2011sparsity}. To make a fair comparison, we consider only the SMV case when comparing our method with STLS though a similar problem can be cast for STLS in the MMV case. In our implementation of OGSBI, we initialize $\alpha_0=\frac{100}{ Var\lbra{\m{y}}}$ and $\m{\alpha}=\frac{1}{MK}\abs{\m{A}^H\m{y}}$. The rest settings are the same as those for OGSBI-SVD in the MMV case.

In our experiment, we consider two DOAs from $63.2^\circ$ and $90.3^\circ$ with $\text{SNR}=20$dB. We consider $r=2^\circ$ and $4^\circ$ for both OGSBI and STLS. The parameter $\lambda$ in (\ref{formu:problem_STLS}) is tuned to our best such that STLS achieves the smallest error. Table \ref{table:comparison_STLS} presents the averaged MSEs and CPU times of STLS and OGSBI over $R=200$ trials. OGSBI obtains more accurate DOA estimations than STLS in both the scenarios with remarkably less computational times. We also note that it is possible to accelerate STLS using state-of-the-art algorithms for CS.

\begin{table}
 \caption{Averaged MSEs and CPU Times of STLS and OGSBI in the SMV Case With Respect to $r$.}
 \centering
\begin{tabular}{l|r|r|r|r}
  \hline\hline
  &\multicolumn{2}{c|}{MSE (dB)} & \multicolumn{2}{c}{Time (sec)}\\
  \hline
   & $r=2^\circ$ & $r=4^\circ$ & $r=2^\circ$ & $r=4^\circ$ \\\hline
  STLS & $-36.5$ & $-36.6$ & $5.31$ & $1.77$ \\
  OGSBI & $-45.1$ & $-43.3$ & $0.098$ & $0.028$ \\
  \hline\hline
\end{tabular}\label{table:comparison_STLS}
\end{table}

\subsection{Sensitivity to Measurement Outliers}
The SVD procedure in OGSBI-SVD is related to the principal component analysis (PCA). As is known that the standard PCA is sensitive to outliers. Even a single corrupted measurement can deteriorate the quality of the approximation. In this subsection we carry out experiments to check whether the proposed OGSBI-SVD is sensitive to measurement outliers due to the SVD. The experimental setup is similar to that in Subsection \ref{sec:ComparisonWithL1SVD} but with $\text{SNR}=\infty$. After acquiring the noiseless measurements, we randomly choose 3 out of the $MT=1600$ measurements, multiply by a constant ratio $\kappa$ and then save as the outliers. Beside the case of no outliers ($\text{ratio}=1$) we consider five other cases where $\kappa$ is set to 5, 10, 20, 50 and 100 respectively. Table \ref{table:outlier} presents our simulation results of the MSEs. It can be seen that the estimation accuracy of OGSBI-SVD can degrade significantly even with about $0.2\%$ measurements being corrupted due to the sensitivity of the SVD.

\begin{table}
 \caption{Averaged MSEs of OGSBI-SVD in the Presence of Outliers. $\kappa$ denotes a ratio by which outliers are augmented.}
 \centering
\begin{tabular}{l|r|r|r|r|r|r}
  \hline\hline
  $\kappa$ & 1 & 5 & 10 & 20 & 50 & 100 \\\hline
  MSE(dB) & $-63.1$ & $-62.9$ & $-47.6$ & $-38.6$ & $-34.6$ & $-32.0$ \\
  \hline\hline
\end{tabular}\label{table:outlier}
\end{table}

Note that the corrupted measurement matrix $\m{Y}$ due to the outliers is a sum of a low-rank matrix (noiseless measurement matrix of rank $K$) and a sparse matrix (outliers). A robust PCA technique has recently been proposed in \cite{candes2011robust} that can recover the original low-rank matrix from the sparse outliers. So, it is possible to combine the robust PCA technique in \cite{candes2011robust} with the proposed OGSBI-SVD to improve its robustness to outliers, which, however, is beyond the scope of this paper.

\section{Conclusion}\label{sec:discussion}

In this paper, we studied the off-grid DOA estimation model firstly proposed in \cite{zhu2011sparsity} for reducing the modeling error due to discretization of a continuous range. We proposed an algorithm based on the off-grid model from a Bayesian perspective that is applicable to both single snapshot and multi-snapshot cases. A subspace-based idea was used to reduce the computational complexity of the signal recovery process and the sensitivity to noise. We illustrated by simulations that the proposed approach outperforms standard CS methods whose performance is limited by the underlying standard on-grid model. It is also more accurate than the algorithm in \cite{zhu2011sparsity} based on the off-grid model. One drawback of the proposed algorithm is its slow speed in the case of a dense sampling grid though a coarser grid can be adopted to obtain an accurate yet fast DOA estimation. A future work is to develop fast versions of our algorithm. After this work, we have shown in \cite{yang2011robustly} that $\ell_1$ optimization also works for the off-grid DOA estimation problem, where performance guarantees are also provided under some conditions.

\section*{Appendix: Derivation of (\ref{formu:solve_beta})}
Denote $\m{\Delta}=\diag\sbra{\m{\beta}}$. Eq. (\ref{formu:solve_beta}) is based on the following two equalities:
\[\begin{split}
&\twon{\m{y}-\sbra{\m{A}+\m{B}\m{\Delta}}\m{\mu}}^2\\
&=\twon{\sbra{\m{y}-\m{A}\m{\mu}} - \m{B}\cdot\diag\sbra{\m{\mu}}\cdot\m{\beta}}^2\\
&=\m{\beta}^T\sbra{\overline{\m{B}^H\m{B}}\odot\m{\mu}\m{\mu}^H}\m{\beta} \\
&\quad- 2\Re\lbra{\diag\sbra{\overline{\m{\mu}}}\m{B}^H\sbra{\m{y}-\m{A}\m{\mu}}}^T\m{\beta}+C_1,\\
&Tr\lbra{\sbra{\m{A}+\m{B}\m{\Delta}}\m{\Sigma}\sbra{\m{A}+\m{B}\m{\Delta}}^H}\\
&=2\Re\lbra{Tr\lbra{\m{B}^H\m{A}\m{\Sigma}\m{\Delta}}} + Tr\lbra{\m{\Delta}\m{\Sigma}\m{\Delta}\m{B}^H\m{B}}+C_2\\
&=2\Re\lbra{\diag\sbra{\m{B}^H\m{A}\m{\Sigma}}}^T\m{\beta} + \m{\beta}^T\sbra{\m{\Sigma}\odot\overline{\m{B}^H\m{B}}}\m{\beta}+C_2
\end{split}\]
where $C_1$, $C_2$ are constants independent of $\m{\beta}$, and the equality \[Tr\lbra{\diag^H\sbra{\m{u}}\m{Q}\cdot\diag\sbra{\m{v}}\cdot\m{R}^T} =\m{u}^H\sbra{\m{Q}\odot\m{R}}\m{v}\]
for vectors $\m{u},\m{v}$ and matrices $\m{Q},\m{R}$ with proper dimensions is used. Note that $\m{\beta}^T\m{S}\m{\beta}\in\bR$ for a positive semi-definite matrix $\m{S}$ with proper dimension and thus
$\m{\beta}^T\m{S}\m{\beta} = \Re\lbra{\m{\beta}^T\m{S}\m{\beta}} = \m{\beta}^T\cdot\Re\m{S}\cdot\m{\beta}$
since $\m{\beta}$ is real-valued.
Then (\ref{formu:solve_beta}) is obtained by observing that both $\overline{\m{B}^H\m{B}}\odot\m{\mu}\m{\mu}^H$ and $\m{\Sigma}\odot\overline{\m{B}^H\m{B}}$ are positive semi-definite.

\bibliographystyle{IEEEtran}


\end{document}